\documentclass[twocolumn,prd,amsmath,amssymb,nofootinbib]{revtex4}
\usepackage{graphicx}

\renewcommand{\vec}[1]{\boldsymbol{\mathrm{#1}}}

\begin{document}

\title{Scalar--Tensor--Vector modified gravity in light of the Planck 2018 data}

\author{J. W. Moffat$^{\star}$ and
V. T. Toth$^\star$\\~\\
{\rm
\footnotesize
$^\star$Perimeter Institute for Theoretical Physics, Waterloo, Ontario N2L 2Y5, Canada}}

\begin{abstract}
The recent data release by the Planck satellite collaboration presents a renewed challenge for modified theories of gravitation. Such theories must be capable of reproducing the observed angular power spectrum of the cosmic microwave background radiation. For modified theories of gravity, an added challenge lies with the fact that standard computational tools do not readily accommodate the features of a theory with a variable gravitational coupling coefficient. An alternative is to use less accurate but more easily modifiable semianalytical approximations to reproduce at least the qualitative features of the angular power spectrum. We extend a calculation that was used previously to demonstrate compatibility between the Scalar--Tensor--Vector--Gravity (STVG) theory, also known by the acronym MOG, and data from the Wilkinson Microwave Anisotropy Probe (WMAP) to show consistency between the theory and the newly released Planck 2018 data. We find that within the limits of this approximation, the theory accurately reproduces the features of the angular power spectrum.
\end{abstract}

\maketitle

\section{Introduction}

Though highly isotropic, the cosmic microwave background (CMB) shows small temperature fluctuations as a function of sky direction. The magnitude of these fluctuations depends on angular size. This location and size of these peaks is an important prediction of the standard model of cosmology, which has been confirmed by increasingly accurate experiments, such as the Boomerang experiment\cite{Jones2006}, the Wilkinson Microwave Anisotropy Probe (WMAP, \cite{Komatsu2008}), and the Planck satellite \cite{Planck2020Paper6}.

The angular power spectrum of the CMB can be calculated in a variety of ways. The preferred method is to use numerical software, such as {\tt CMBFAST}\cite{Seljak1996}. Unfortunately, such software packages cannot easily be adapted for use with a variable-$G$ theory of gravitation, such as Scalar--Tensor--Vector--Gravity (STVG \cite{Moffat2005}) theory, also known as MOdified Gravity (MOG).

There are alternative methods of calculation, which, though somewhat less accurate, nonetheless capture the essential qualitative features of the CMB angular power spectrum. The advantage of such calculations is that the physics is transparent, not obscured by ``black box'' computer code, and that the calculations can be adapted with relative ease to accommodate a different theory. One such method is the semianalytical approximation presented in Ref.~\cite{Mukhanov2005}.

We have, in fact, used this approximation in the past showing agreement between the predictions of the MOG theory and the WMAP results \cite{Moffat2007c,Moffat2013a}. In light of the recently published Planck 2018 results, it is important to revisit and refine this computation, and also extend it to high values of the multipole index $\ell$.

We begin in Sec.~\ref{sec:MOG} with a brief introduction to the MOG theory and its acceleration law, which gives rise to the theory's effective gravitational coupling parameter. In Sec.~\ref{sec:CMB} we introduce the angular power spectrum and its semianalytical approximation, as presented in Ref.~\cite{Mukhanov2005} (which the interested reader is advised to consult for details.) We adapt the calculation to the MOG theory and show results.
We conclude by presenting our discussion and conclusions in \ref{sec:END}.

\section{The MOG theory}
\label{sec:MOG}

Our MOG modified gravity theory, also known as Scalar--Tensor--Vector--Gravity (STVG \cite{Moffat2005}), is a relativistic theory of gravitation based on an action principle. In addition to the metrical field of gravitation, the theory introduces a repulsive vector field of finite range. The gravitational constant and the vector field range (mass) parameter are promoted to dynamical (massless) scalar fields. Within the range of the vector field, the theory replicates Newtonian gravitation; outside this range, in the absence of the repulsive force, gravitation is stronger. By this feature, the theory successfully accounts for galaxy rotation curves \cite{MoffatRahvar2013,MOG2015,GreenMoffat2019,DavariRahvar2020,Moffat2020a}, the matter power spectrum \cite{Moffat2013a}, while also remaining consistent with recent gravitational wave data \cite{MOG2017a}.

In the weak field, low-velocity regime, the MOG theory yields a simple gravitational acceleration law \cite{Moffat2007e}. For a point source of gravitation characterized by mass $M$ at the origin, the gravitational acceleration at position $\vec{r}$ is given by
\begin{align}
\ddot{\vec{r}}=-\frac{G_{\rm eff}M}{r^3}\vec{r},
\end{align}
with
\begin{align}
G_{\rm eff}=G_N\left[1+\alpha-\alpha(1+\mu r)e{^{-\mu r}}\right],
\end{align}
where $G_N$ is Newton's constant of gravitation, the dimensionless quantity $\alpha=(G-G_N)/G_N$ (i.e., $G=(1+\alpha)G_N$) characterizes the difference between the theory's variable gravitational coupling coefficient $G$ and $G_N$, and $\mu$ is the mass of the vector field.

In an approximately homogeneous and isotropic universe, $\alpha$ and $\mu$ can be taken as constants. Consequently, at distance scales characterized by $\mu r\gg 1$, $G_{\rm eff}\sim G_N(1+\alpha)$ can be treated as constant as well.

The Friedmann equations that describe a homogeneous and isotropic universe remain valid in the MOG theory \cite{Moffat2007c,Toth2010b,Moffat2013a,Moffat2014a}, with only trivial modifications. Not surprising, given that these equations can also be heuristically derived from the Newtonian theory \cite{Mukhanov2005,Weinberg2008}. The equations read (using $c=1$):
\begin{align}
\frac{\dot{a}^2}{a^2}+\frac{k}{a^2}&=\frac{8\pi G_{\rm eff}\rho}{3}+\frac{\Lambda}{3},\\
\frac{\ddot{a}}{a}&=-\frac{4\pi G_{\rm eff}}{3}\left(\rho+3p\right)+\frac{\Lambda}{3}.
\end{align}
The critical density, characterized by $k=0$, $\Lambda=0$, is given by
\begin{align}
\rho_{\rm crit}^{\rm MOG}=\frac{3H^2}{8\pi G_{\rm eff}},
\end{align}
where $H=\dot{a}/a$.

Note the presence of a factor of $1/(1+\alpha)$ in this definition of $\rho_{\rm crit}$. Consequently, for a given baryon density $\rho_b$, the corresponding density parameter $\Omega_b$ is inflated by this same factor:
\begin{align}
\Omega_b^{\rm MOG}=\frac{\rho_b}{\rho_{\rm crit}^{\rm MOG}}=
(1+\alpha)\frac{8\pi G_N\rho_b}{3H^2}=(1+\alpha)\Omega_b.
\end{align}
Often, in cosmological calculations, $\Omega_b$ is used to represent the baryon density in equations describing both gravitational and nongravitational interactions. Clearly, this convenience is lost in the case of modified gravity, in the presence of the $(1+\alpha)$ factor, which applies only to gravitational interactions.

\section{Modeling the Cosmic Microwave Background (CMB)}
\label{sec:CMB}

It is surprisingly difficult to analyze high quality Cosmic Microwave Background (CMB) data sets from the perspective of a modified gravity theory, such as MOG. The main reason for this difficulty lies with the fact that, as we alluded to above, the dimensionless density parameter $\Omega_b$ is used to represent baryonic matter in calculations that involve gravity as well as calculations that represent nongravitational physics.

Why is this a problem? Consider the definition:
\begin{align}
\Omega_b=\frac{\rho_b}{\rho_{\rm crit}}=\frac{8\pi G\rho_b}{3H^2}.
\end{align}
In the standard theory, this expression will suffice. But what about a theory, such as MOG, with a variable gravitational coefficient $G=G_{\rm eff}=(1+\alpha)G_N$? Clearly, when the context is gravitational, the product $G\rho_b$ accurately reflects the gravitational contribution of baryonic matter. But when, e.g., the pressure of the medium is considered, $\Omega_b$ is not supposed to be scaled in this manner (pressure does not increase just because gravitation is stronger).

Disentangling these issues in computer codes that have been in use for years or decades, written or rewritten by multiple authors, perhaps even machine translated from one programming language to another (e.g., from {\rm FORTRAN} to C) is a daunting task.

Without access to a standard suite of computer programs that can reliably and provably deal with a variable-$G$ modified theory of gravity, we opted for another approach: use a semianalytical approximation that is sufficiently accurate to reproduce the key qualitative features of the CMB angular power spectrum, and perhaps even allow us to make some cautious predictions.

Such an approximation method was published by Mukhanov \cite{Mukhanov2005}. We previously used this approximation method in the context of WMAP results, showing that MOG indeed fits the angular power spectrum well. In light of the recent release of Planck 2018 data, we found it imperative to revisit and, if necessary, refine this calculation and compare Planck results against MOG predictions.

\subsection{Semi-analytical estimation of CMB anisotropies}

The general expression for the cosmic mean of the CMB temperature autocorrelation function, expressed in terms of multipoles $C_\ell$ (with the monopole and dipole components, $\ell=0,1$, excluded), can be written as (see Eq.~(9.38) in \cite{Mukhanov2005} and the discussion therein for details):
\begin{align}
C_\ell=\frac{2}{\pi}\int\Bigg|\bigg(\Phi_k(\eta_r)+{}&\frac{\delta_k(\eta_r)}{4}\bigg)j_\ell(k\eta_0)\nonumber\\
&{}-\frac{3\delta_k'(\eta_r)}{4k}\frac{dj_\ell(k\eta_0)}{d(k\eta_0)}\Bigg|^2k^2dk,
\end{align}
where $\Phi_k$ is the Fourier-decomposition of the gravitational potential $\Phi$ with respect to wavenumber $k$,
$\eta_r$ is the conformal time at recombination and $\eta_0$ corresponds to the present time.
The quantity $\delta$ is the fractional energy fluctuation of radiation, defined using the $00$-component of the radiation energy-momentum tensor before recombination as $T_0^0=\epsilon(1+\delta)$ where $\epsilon$ is the radiation energy density; $\delta'$ is the derivative with respect to conformal time and and $j_\ell$ are the spherical Bessel functions.

For $k\eta_r\ll 1$, $\delta_k(\eta_r)\simeq -\frac{8}{3}\Phi_k(\eta_r)$, $\delta_k'(\eta_r)\simeq 0$, hence we find that for $\ell\ll 200$, $\ell(\ell+1)C_\ell\simeq{\rm const.}$ This observation is valid both in the standard $\Lambda$-CDM cosmology and the MOG theory, leaving us, for low $\ell$, with
\begin{align}
C_l=\frac{2}{\pi}\int\left|\frac{1}{3}\Phi_k(\eta_r)j_l(k\eta_0)\right|^2k^2dk.
\end{align}
If $|\Phi_k|^2=(9/10)^2B/k^3$ (the extra factor $9/10$ corresponding to a drop of the potential on superhorizon scales after matter-radiation equality), we get
\begin{align}
C_l=\frac{18B}{100\pi}\int j_l(k\eta_0)^2k^{-1}dk.
\end{align}
Let $s=k\eta_0$, $ds/dk=\eta_0$, and then,
\begin{align}
C_l=\frac{18B}{100\pi}\int j_l(s)^2s^{-1}ds=\frac{9B}{100\pi l(l+1)}.
\end{align}
For large $\ell$, still following Ref.~\cite{Mukhanov2005}), we can then write
\begin{align}
\frac{\ell(\ell+1)C_\ell}{[\ell(\ell+1)C_{\ell}]_{{\rm low}~\ell}}=\frac{100}{9}(O+N),
\end{align}
where we split the eventual solution into an oscillatory ($O$) and non-oscillatory ($N$) part.

Using approximations of the Bessel function and other suitable numerical representations, we find the following expression for the oscillatory part \cite{Mukhanov2005}:
\begin{align}
O=&e^{-(l/l_s)^2}\sqrt{\frac{\pi}{\bar\rho l}}\nonumber\\
&{}\times\left[A_1\cos{\left(\bar\rho l+\frac{\pi}{4}\right)}+A_2\cos{\left(2\bar\rho l+\frac{\pi}{4}\right)}\right],
\end{align}
where
\begin{align}
A_1=0.1\xi\frac{(P-0.78)^2-4.3}{(1+\xi)^{1/4}}e^{\frac{1}{2}(l_s^{-2}-l_f^{-2})l^2},
\end{align}
and
\begin{align}
A_2=0.14\frac{(0.5+0.36P)^2}{(1+\xi)^{1/2}}.
\end{align}
The non-oscillatory part, in turn, is split into a sum:
\begin{align}
N=N_1+N_2+N_3,
\end{align}
where
\begin{align}
N_1=0.063\xi^2\frac{[P-0.22(l/l_f)^{0.3}-2.6]^2}{1+0.65(l/l_f)^{1.4}}e^{-(l/l_f)^2},
\end{align}
\begin{align}
N_2=\frac{0.037}{(1+\xi)^{1/2}}\frac{[P-0.22(l/l_s)^{0.3}+1.7]^2}{1+0.65(l/l_s)^{1.4}}e^{-(l/l_s)^2},
\end{align}
\begin{align}
N_3=\frac{0.033}{(1+\xi)^{3/2}}\frac{[P-0.5(l/l_s)^{0.55}+2.2]^2}{1+2(l/l_s)^2}e^{-(l/l_s)^2}.
\end{align}

The parameters that occur in these expressions are as follows. First, the baryon density parameter:
\begin{align}
\xi=17\left(\Omega_bh_{75}^2\right),
\label{eq:xi}
\end{align}
where $\Omega_b\simeq 0.035$ is the baryon content of the universe at present relative to the critical density, and $h_{75}=H_0/(75~\mathrm{km/s/Mpc})$ with $H_0$ being the Hubble parameter at the present epoch. The growth term of the transfer function is represented by
\begin{align}
P=\ln{\frac{\Omega_m^{-0.09}l}{200\sqrt{\Omega_mh_{75}^2}}},
\end{align}
where $\Omega_m\simeq 0.3$ is the total matter content (baryonic matter, neutrinos, and cold dark matter). The free-streaming and Silk damping scales are determined, respectively, by
\begin{align}
l_f=1600\left[1+7.8\times 10^{-2}\left(\Omega_mh_{75}^2\right)^{-1}\right]^{1/2}\Omega_m^{0.09},
\label{eq:lf}
\end{align}
\begin{align}
l_s=\frac{0.7l_f}{\sqrt{\frac{1+0.56\xi}{1+\xi}+\frac{0.8}{\xi(1+\xi)}\frac{\left(\Omega_mh_{75}^2\right)^{1/2}}{\left[1+\left(1+\frac{100}{7.8}\Omega_mh_{75}^2\right)^{-1/2}\right]^2}}}.
\end{align}
Lastly, the location of the acoustic peaks is determined by the parameter\footnote{Note the slight changes in the coefficients in (\ref{eq:lf}) and (\ref{eq:rho}) compared to the value published in \cite{Mukhanov2005}. We used best fit values for these coefficients from Mukhanov's approximation using the Planck collaboration's best estimates for the parameters of the standard $\Lambda$CDM cosmology.}.
\begin{align}
\bar\rho=0.015(1+0.13\xi)^{-1}(\Omega_mh_{75}^{3.1})^{0.16}.
\label{eq:rho}
\end{align}

Finally, we note that the calculated result for $C_\ell$ assumes scale invariance. For small deviations from scale invariance characterized as usual by the parameter $n_s$ (with $|n_s-1|\ll 1$), the result is scaled:
\begin{align}
C_\ell\to \ell^{n_s-1}C_\ell.
\end{align}

\urldef\esawiki\url{https://wiki.cosmos.esa.int/planck-legacy-archive/index.php/CMB_spectrum_%26_Likelihood_Code}

\begin{figure*}[t]
\includegraphics[width=0.49\linewidth]{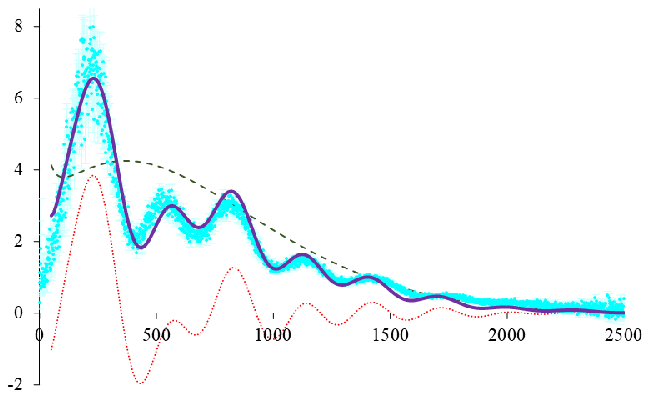}~\includegraphics[width=0.49\linewidth]{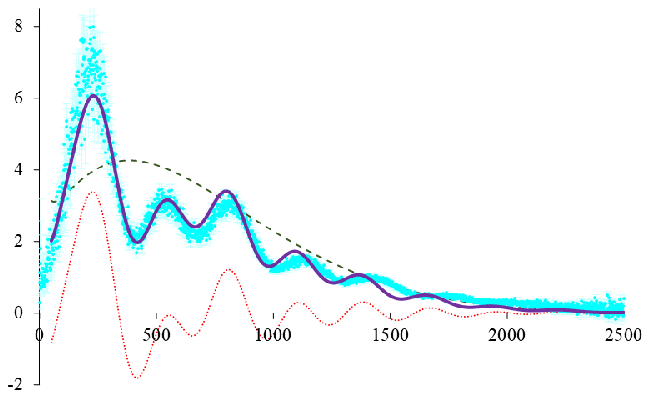}
\includegraphics[width=0.49\linewidth]{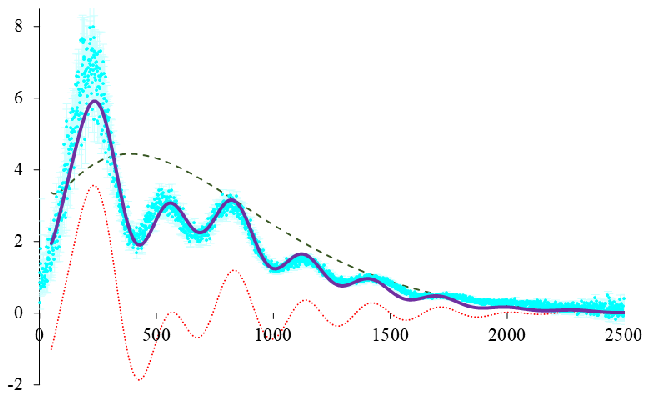}~\includegraphics[width=0.49\linewidth]{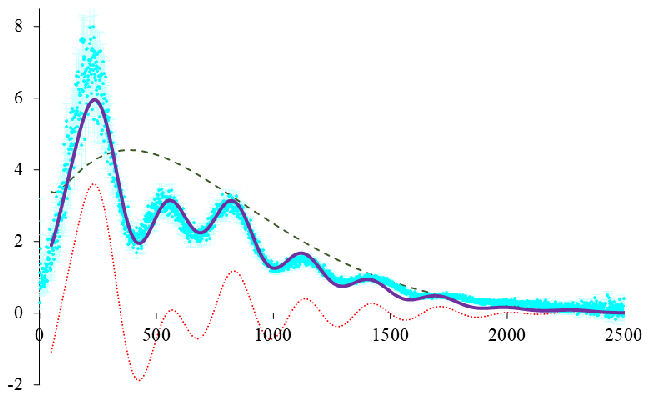}
\caption{\label{fig:CMB}Mukhanov's approximation of the angular power spectrum in light of Planck 2018 data. Thick blue line: Mukhanov's approximation as a sum of an oscillatory part (thin dotted red line) and nonoscillatory part (dashed green line). Planck 2018 data are shown in light blue with vertical error bars. Top row: standard cosmology, with nominal parameters (left) and least squares fitted values for $\Omega_b$ and $n_s$ (right). Bottom row: The MOG theory, with $\Omega_b$, $\alpha$ and $n_s$ fitted (left) and with the same 3-parameter fit but setting $H_0=73$~km/s/Mpc fitted (right).}
\end{figure*}

The quality of this approximation is demonstrated in Fig.~\ref{fig:CMB} (top left), which shows the estimated angular power spectrum using nominal parameters ($H_0\sim 67.4$~km/s/Mpc, $h^2\Omega_b=0.0224$, $\Omega_m=0.315$, $n_s=0.965$ with spatially flat cosmology, $\Omega_\Lambda=1-\Omega_m$), against Planck 2018 data\footnote{For important explanations see \esawiki.}
from \url{http://pla.esac.esa.int/pla/#cosmology}.

The quality of this fit improves significantly if we allow some of the parameters to vary. For instance, using a simple least squares fit, we obtain $h^2\Omega_b=0.0187$, $n_s=0.965$ (see Fig.~\ref{fig:CMB} top right).

\subsection{The MOG CMB spectrum}

What are the key differences between the MOG theory and standard cosmology?

In the standard $\Lambda$CDM model in the early universe, there are two main sources of gravitation: baryonic matter and collisionless cold dark matter (CDM). The distribution of matter in the universe is still largely homogeneous, and the gravitational field is determined by the sum $\Omega_m=\Omega_b+\Omega_{\rm DM}$.

In the MOG theory, $\Omega_{\rm DM}$ is, of course, absent. However, the gravitational coupling parameter is no longer Newton's constant. In the late time universe, we expect the gravitational coupling parameter to vary from region to region (an essential feature of the MOG theory that accounts for its ability to model phenomena such as galaxy rotation curves successfully.) In the early, mostly homogeneous universe, we expect little variation in the value of $G$; however, $G\ne G_N$.

This means that gravitational interactions are ``enhanced'' by the factor $1+\alpha$, defined by the relationship $G=G_{\rm eff}=(1+\alpha)G_N$. When computing results such as the angular power spectrum, this must be taken into account.

This actually leads to a fairly simple prescription. In the formulation presented in the previous subsection, the density parameter for matter, $\Omega_m$, must be replaced by $(1+\alpha)\Omega_b$.

These changes are, of course, trivial. $\Omega_b$ appears only in Eq.~(\ref{eq:xi}). For otherwise identical parameterization, we expect identical results.

Instead, we opted to relax the parameter space further as we investigate the MOG solution. Fig.~\ref{fig:CMB}, bottom left, was obtained by fitting the values of $h^2\Omega_b=0.0197$, $\alpha=5.27$ and $n_s=0.951$.

As we explored the parameter space, it became evident that there is significant degeneracy with respect to the value of $H_0$. Fig.~\ref{fig:CMB} (lower right) shows another fit, after setting $H_0=73$~km/s/Mpc, resulting in $h^2\Omega_b=0.0199$, $\alpha=4.75$ and $n_s=0.949$. We believe that this degeneracy demonstrates the limit of the Mukhanov approximation.

\section{Conclusions}
\label{sec:END}

Recently, the Planck collaboration released a data set characterizing the cosmic microwave background's angular power spectrum in more detail than anything previously published. This data release raises the bar for modified theories of gravitation that compete with the standard $\Lambda$CDM model as potentially viable representations of the evolution of, and structure formation in the universe.

We investigated in particular the behavior of Scalar--Tensor--Vector--Gravity, also known by the acronym MOG, in the light of these new data. A key feature of the MOG theory is the presence of a variable gravitational coupling coefficient, which makes the task of adapting existing numerical models of the CMB or structure formation difficult. Large numerical code bases that are opaque and often use the dimensionless density parameters $\Omega_b$, $\Omega_m$, etc., to model both gravitational and nongravitational interactions, cannot be easily modified.

Instead, in this paper, we revived a model that we first employed in the wake of the WMAP data release. Extending the calculations to higher multipoles (up to $\ell=2500$) we were able to demonstrate that the MOG theory correctly reproduces the qualitative features of the CMB, and that within the limits of the approximation, also produces good quantitative fits. At the same time, we also saw the limitations of the method, notable among them a degeneracy with respect to $H_0$. This leads us to conclude that, for instance, to decide whether or not the MOG theory can offer a better resolution to the Hubble tension (see \cite{Hubble2021review} for an up-to-date review), more sophisticated methods will be required.

\acknowledgements

This research was supported in part by Perimeter Institute for Theoretical Physics. Research at Perimeter Institute is supported by the Government of Canada through the Department of Innovation, Science and Economic Development Canada and by the Province of Ontario through the Ministry of Research, Innovation and Science.
VTT acknowledges the generous support of Plamen Vasilev and other Patreon patrons.

\bibliography{refs}
\bibliographystyle{apsrev}

\end{document}